\documentclass[preprint,12pt]{elsarticle}

\usepackage[T1]{fontenc}
\usepackage{newtxtext,newtxmath}

\usepackage{amsmath}
\usepackage{bm}

\usepackage{algorithm}
\usepackage[noend]{algpseudocode}

\usepackage{graphicx}
\usepackage{subcaption}
\usepackage{booktabs}
\usepackage{array}
\usepackage{threeparttable}
\usepackage{makecell}
\usepackage{tabularx}
\usepackage{multirow}
\usepackage{tablefootnote}

\usepackage{pgfplots}
\pgfplotsset{compat=1.18}
\usepgfplotslibrary{fillbetween}
\usetikzlibrary{patterns,arrows.meta}

\usepackage{geometry}
\usepackage{lineno}
\usepackage{natbib}

\usepackage[hidelinks]{hyperref}

\AtBeginDocument{%
  \pdfstringdefDisableCommands{%
    \def\corref#1{}%
    \def\cortext#1{}%
  }%
}

\makeatletter
\def\ps@pprintTitle{%
  \let\@oddhead\@empty
  \let\@evenhead\@empty
  \let\@oddfoot\@empty
  \let\@evenfoot\@oddfoot
}
\makeatother

\DeclareUnicodeCharacter{202F}{\,} 
\DeclareUnicodeCharacter{2011}{-}  
\usepackage[hidelinks]{hyperref}

\begin{document}
\begin{frontmatter}

\title{PAPPL: Personalized AI-Powered Progressive Learning Platform}

\author[1]{Shayan Bafandkar}
\author[1]{Sungyong Chung}
\author[2]{Homa Khosravian}
\author[1]{Alireza Talebpour\corref{cor1}}
\cortext[cor1]{Corresponding author: \texttt{ataleb@illinois.edu}}
\address[1]{The Grainger College of Engineering, Department of Civil and Environmental Engineering, University of Illinois at Urbana-Champaign, 205 N Matthews, Urbana, IL 61801}
\address[2]{National Science Foundation, 2415 Eisenhower Ave, Alexandria, VA 22314}

\begin{abstract}
Engineering education has historically been constrained by rigid, standardized framework, often neglecting students' diverse learning needs and interests. While significant advancements have been made in online and personalized education within K–12 and foundational sciences, engineering education at both undergraduate and graduate levels continues to lag in adopting similar innovations. Traditional evaluation methods, such as exams and homework assignments, frequently overlook individual student requirements, impeding personalized educational experiences. To address these limitations, this paper introduces the Personalized AI-Powered Progressive Learning (PAPPL) platform, an advanced Intelligent Tutoring System (ITS) designed specifically for engineering education. It highlights the development of a scalable, data-driven tutoring environment leveraging cutting-edge AI technology to enhance personalized learning across diverse academic disciplines, particularly in STEM fields. PAPPL integrates core ITS components including the expert module, student module, tutor module, and user interface, and utilizes GPT-4o, a sophisticated large language model (LLM), to deliver context-sensitive and pedagogically sound hints based on students’ interactions. The system uniquely records student attempts, detects recurring misconceptions, and generates progressively targeted feedback, providing personalized assistance that adapts dynamically to each student's learning profile. Additionally, PAPPL offers instructors detailed analytics, empowering evidence-based adjustments to teaching strategies. This study provides a fundamental framework for the progression of Generative ITSs scalable to all education levels, delivering important perspectives on personalized progressive learning and the wider possibilities of Generative AI in the field of education.
\end{abstract}


\end{frontmatter}




\section{Introduction}
Engineering education has long been constrained by traditional, rigid structures that prioritize standardized curricula and uniform pacing over flexibility and individual learning styles. As a result, many students find it challenging to explore cross-disciplinary interests, progress at their own speed, or adapt their education to evolving career demands. While numerous initiatives have successfully advanced online learning in K–12 and foundational science education, including mathematics~\cite{barnes2010automatic,levonian2025designing}, life science~\cite{joyner2015inquiry}, computer science~\cite{piech2015autonomously}, and electrical engineering~\cite{graesser2018electronix}, there remains a significant gap in similar efforts tailored specifically to engineering.

Traditionally, student evaluation in engineering education relies heavily on exams, homework assignments, and quizzes, a practice still widespread in many engineering courses at the nation's top institutions. However, these conventional evaluation methods often overlook students' diverse learning needs and may inadvertently impede their academic progress~\cite{trifoni2011exam, grassi2011technologies}.
Addressing this limitation and with the emergence of Artificial Intelligence (AI) in the form of foundational models, the focus has been shifted to developing Intelligent Tutoring Systems (ITSs). These advanced learning platforms, powered by large language models (LLMs), are designed to support both instructors and students across various disciplines and educational levels without requiring extensive training for different subjects.

Another motivation behind developing ITSs is the increasing demand for personalized education tailored to individual students' backgrounds and needs~\cite{ciloglugil2010adaptive}. A good way to ensure that ITS responses are personalized is to keep them interactive by asking questions that guide students through the learning process and in areas where they might experience difficulties. Several ITSs have explored question-based feedback to enhance student learning. AutoTutor~\cite{1532370} employs mixed-initiative natural language dialogues to simulate a human tutor, posing open-ended questions that encourage active reasoning. Korbit~\cite{serban2020large} scales this idea with a large-scale, mixed-interface system that dynamically generates follow-up questions based on misconceptions identified through machine learning. SMILE~\cite{buckner2014integrating} shifts the question-generation responsibility to students themselves, promoting metacognitive skills by having them create and evaluate peer-generated questions. Betty's Brain~\cite{leelawong2008betty} employs a teachable agent model that encourages learners to reflect through feedback cycles triggered by their agent's performance. Moreover, the ItsSQL~\cite{aguirrereid2023itssql} system provides targeted feedback on structured queries, helping students identify syntax and logic errors. Despite their diversity, these systems often treat feedback as either static or based solely on students' current input.

In general, ITSs have four core components, described as expert module, student module, tutor module, and user interface~\cite{ma2017systematic}. Each component fulfills a distinct role, from encoding subject matter knowledge and tracking the learner's understanding to selecting instructional strategies and facilitating human-computer interaction, collectively enabling personalized instruction. To the best of our knowledge, previous ITS implementations rarely achieved a fully integrated balance of all four components. In many cases, the focus of ITS development efforts has been on a particular component while simplifying or neglecting others, and no single ITS architecture encompassed the complete range of tutoring strategies, given the fragmented nature of past designs~\cite{nwana1990overview}. Consequently, many ITS efforts were confined to narrow, well-structured domains, limiting their generalizability beyond those constrained contexts~\cite{liu2024advancing}. Another gap is related to students' learning assessment. An effective ITS should capture students' progress, assess their learning, and generate appropriate responses accordingly to improve their learning~\cite{ma2017systematic}. These gaps highlight the need for a more unified and scalable approach to intelligent tutoring.

This paper introduces the Personalized AI-Powered Progressive Learning (PAPPL) platform, a robust educational tool providing personalized AI-generated hints based on student interactions, which can significantly enhance the learning experience. PAPPL aims to tailor its feedback uniquely for each student by storing previous responses, identifying patterns of frequent errors, and personalizing responses accordingly. The platform leverages sophisticated Generative Pre-trained Transformer (GPT) functionalities that build context-aware prompts from comprehensive student interaction histories and instructor-provided contextual descriptions, ensuring feedback is relevant, personalized, and pedagogically sound. By integrating cutting-edge technologies, modular course design, and a broad range of applications, PAPPL can offer a richer, adaptive learning experience, empowering students to take control of their educational journey.

Accordingly, the main contributions of this study are twofold: (1) while previous studies often do not provide dynamic feedback or ignore students' interaction history, our PAPPL system introduces a novel multi-layered feedback core that synthesizes question context, students' prior misconceptions, instructors' explanations, and related question history to generate personalized guiding question-based feedback using an AI core; and (2) it creates a versatile platform applicable to various subjects across all levels and unlike many systems that either reveal answers too soon or lack adaptability, PAPPL is equipped with sophisticated multi-layered prompts that, along with appropriate temperature value adjustments, can ensure pedagogically sound responses that are sensitive to the students' learning process and evolve over time. All in all, PAPPL offers new insights into how temporally-aware, context-rich messages can better support students in their learning process.

The remainder of this paper is organized as follows: We present an overview of ITSs in the next section. Following that, we describe the details of the PAPPL platform. A detailed experimental study involving several students from the transportation groups at the University of Illinois at Urbana-Champaign (UIUC) and George Washington University (GWU) is then presented. This section is followed by a comprehensive analysis of the effectiveness of PAPPL. Finally, the paper concludes with summary remarks and outlines future research needs.

\section{Background}
ITSs emerged in the 1970s from Computer-Aided Instruction (CAI) research~\cite{martin1996constraint} and consist of four core components~\cite{ma2017systematic}, as described below:
\begin{itemize}
    \item The expert module is the core of the ITS and contains all the knowledge that is intended to be taught to the students, including question banks, rules, concepts, and logical statements~\cite{woolf2008intelligent}. This information is utilized (by other components) to assess students learning and track their progress towards the goal of the ITS development effort~\cite{salman2013use,sharma2014survey}. The most challenging aspect of designing the expert module is capturing all the possible variations in students' solutions~\cite{williams2016axis}.
    \item The student module is responsible for capturing the changes in students' learning and tracking their progress~\cite{ma2017systematic}. In other words, this component is responsible for capturing students' weaknesses, mistakes, and progress and feeding that information to other components to guide the learning process~\cite{ma2017systematic}, which can be done based on information as simple as the number of correct answers and as complicated as data structures. Therefore, several models have been utilized for this component, including overlay models~\cite{stansfield1980wumpus}, model tracing~\cite{aleven2002metacognitive, koedinger1997city}, Expectation and Misconception Tailoring (EMT)~\cite{wolfe2013autotutor}, Constraint-Based Modeling (CBM)~\cite{mitrovic2007constraint,mitrovic2012fifteen}, and Bayesian Network Modeling~\cite{vanlehn2005andes}.
    \item The tutor or pedagogical model is responsible for interpreting the data from the student module and providing individualized guidance to enhance students' learning~\cite{ohlsson1996errors}. This is particularly challenging as selecting the correct action and providing the appropriate content to help students depend on the history of student actions. In many cases, simple rule-based cases may not provide a reasonable learning experience. This becomes particularly challenging when dealing with higher education and complex problems as opposed to K-12 education.
    \item The user interface acts as a link between the ITS and the student, collects input from the student, and offers feedback in return. To date, many user interfaces have been developed, including computer-based~\cite{aleven2002metacognitive,ottmar2015interactive}, handwritten-based~\cite{rodrigues2024math}, mobile-based~\cite{brown2006mobile}, virtual reality (VR)-based~\cite{lecorre2012chrysaor,perez2017electrical,vannaprathip2025sdmentor}, and augmented reality (AR)~\cite{ates2025ar,ahuja2025disabled}.
\end{itemize}

Note that not all ITSs have all four components~\cite{conati2009new}, and in many cases, the focus of the efforts has been on a particular component, utilizing existing developments for other components. For instance,~\citet{piech2015autonomously} just focused on the tutor module and compared different hint generation algorithms,~\citet{aleven2002metacognitive,ritter2007cognitive} tried to utilize cognitive tutors for hint generation. These approaches are generally based on educational theories and can be a reliable methodology for hint generation~\cite{piech2015autonomously}. However, to use them on higher education problems, they require a clear cognitive model that makes their implementation extremely hard. In another study,~\citet{barnes2008hint} proposed the Hint Factory that utilized a Markov Decision Process (MDP) formulation to generate the hints for the students.~\citet{fenza2017adaptive} proposed an adaptive tutor module based on reinforcement learning. Their model learns by observing the behavior of human tutors and then, based on an assessment of students' learning, can adapt the hint generation process. Unfortunately, these studies do not offer a scalable approach to designing a tutor module for various types of questions in different fields of study.

In our proposed platform, we have integrated the expert module, student module, and tutor module using a foundational model presenting a promising solution to several longstanding challenges in ITSs, as new foundation models like ChatGPT have shown great understanding capacity in diverse fields~\cite{qin2025tool}. Utilizing advanced LLMs, as transformer-based language models~\cite{vaswani2017attention} trained on immense amounts of data and capable of providing impressive results for various tasks, significantly reduces the need for extensive trial-and-error in learning and cognitive modeling. This is especially relevant compared to statistical or deep neural models, which often face performance and stability challenges~\cite{akkaya2019rubiks,mnih2013atari,pomerleau1988alvinn}. Moreover, LLMs have a much higher potential in tracking users' learning progress and predicting learning outcomes, which enables them to outperform traditional knowledge-tracing methods~\cite{zhang2024predicting}. As a result, LLMs have served an important role in developing new ITSs. For instance,~\citet{venugopalan2025caregiver} explored experiments in prompt engineering using the open-source Llama 3 LLM and provided insights into the integration of LLMs with existing tutoring system instructional models and principles, supporting hybrid tutoring settings through conversational assistance.~\citet{borchers2025llm_adaptivity} introduced a reproducible approach to assess the instructional adaptivity and fidelity of LLMs by evaluating models like Llama3-8B, Llama3-70B, and GPT-4o. Their results indicated that GPT-4o is highly aligned with given instructions and effectively prompts learners with open-ended questions to evaluate their knowledge.~\citet{modran2024} introduced an intelligent chatbot tutoring system that combines Retrieval Augmented Generation (RAG) with a tailored LLM to provide personalized and contextually appropriate assistance for university students, thereby improving their learning experience in higher education.~\citet{llm_tutoring_social_skills} presented a framework that enables instructors to work alongside LLMs to create realistic scenarios for social skills training, thereby enhancing the dynamic development of tutoring content.

Despite all the benefits that LLM can bring to ITSs, a significant challenge remains in their integration.~\citet{letourneau2025systematic} examined the impact of ITSs on the learning and performance of K-12 students, highlighting generally positive outcomes but also noting the need for more extensive research on ethical aspects of using AI for teaching. Moreover,~\citet{tithi2025promise} and~\citet{levonian2025safe} evaluated LLMs' ability to generate explanatory hints for logic problems, finding that while LLMs can augment tutoring systems' ability to generate dynamic feedback, additional modifications are necessary to prevent hallucinations and ensure accuracy and pedagogical appropriateness. PAPPL embeds instructor notes and prior misconceptions directly into each feedback loop, which, along with detailed prompt engineering and LLM temperature value alignment, ensures pedagogical soundness while also allowing for dynamic responses that can be enhanced over time.

We have specifically proposed the use of GPT-4o in this study as the core AI engine for the ITS. Trained on extensive and diverse datasets, GPT-4o can dynamically interpret nuanced student inputs, identify misconceptions through iterative interactions, and generate context-sensitive, adaptive hints without requiring additional learning. Moreover, GPT-4o features advanced reasoning capabilities~\cite{openai2023gpt4}, enabling it to tackle complex, layered problems commonly encountered in higher-level education. Additionally, recent innovations that merge GPT-4o with various intelligent tutoring systems (ITSs), including conversational tutoring systems~\cite{schmucker2024ruffle, ahmed2023chatgpt}, context generating systems~\cite{dan2023educhat}, and script generating systems~\cite{abu-rasheed2024supporting, liu2024advancing}, have demonstrated the capabilities of ITSs integrated with GPT-4o.

It is important to note that while the PAPPL platform design utilizes GPT-4o, it is flexible to utilize more advanced future models (if they emerge). This integration of the main components of the ITS using a foundational model facilitates the development process, enhances adaptability, and provides personalized learning experiences across varied fields. Our study is the first to develop a personalized AI-powered learning platform that progressively generates tailored responses for users learning from each user's attempts and can be applied across diverse academic disciplines, including STEM fields.

\section{PAPPL Framework}
\begin{figure}[!ht]
  \centering
  \includegraphics[width=0.8\textwidth, angle=270]{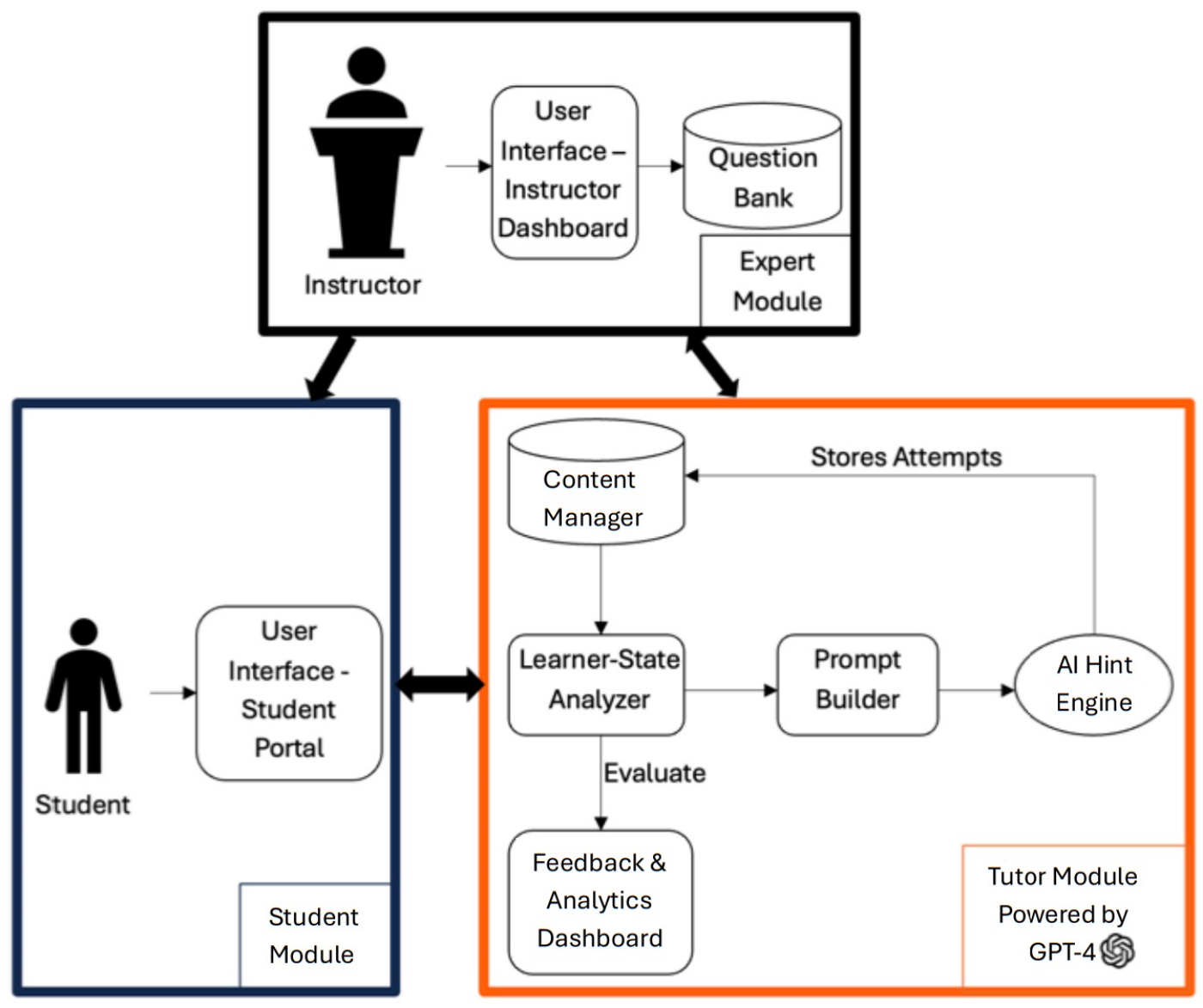}
  \caption{Component Design for PAPPL}
  \label{fig:pappl_arch}
\end{figure}

The PAPPL platform is conceived as a scalable, data-driven tutoring environment that delivers individualized feedback through large-language-model (LLM) services, such as GPT-4o. Its core workflow records each learner’s attempts, detects recurring misconceptions, and synthesizes contextual prompts that ensure eliciting pedagogically sound hints from the AI core. Figure~\ref{fig:pappl_arch} illustrates the information flow among core components:
\begin{itemize}
    \item Frontend: \emph{User Interface Layer}
    \item Backend: \emph{Content Manager}, \emph{Learner-State Analyzer}, \emph{AI Hint Engine}, and the \emph{Feedback \& Analytics Dashboard}
\end{itemize}

Together, these components create a responsive environment where students receive context-aware hints while instructors gain actionable insights into class progress.

\subsection{User Interface Layer}
\begin{figure}[!ht]
    \centering
    \begin{subfigure}[b]{0.45\textwidth}
        \centering
        \includegraphics[width=0.25\textheight, angle=270]{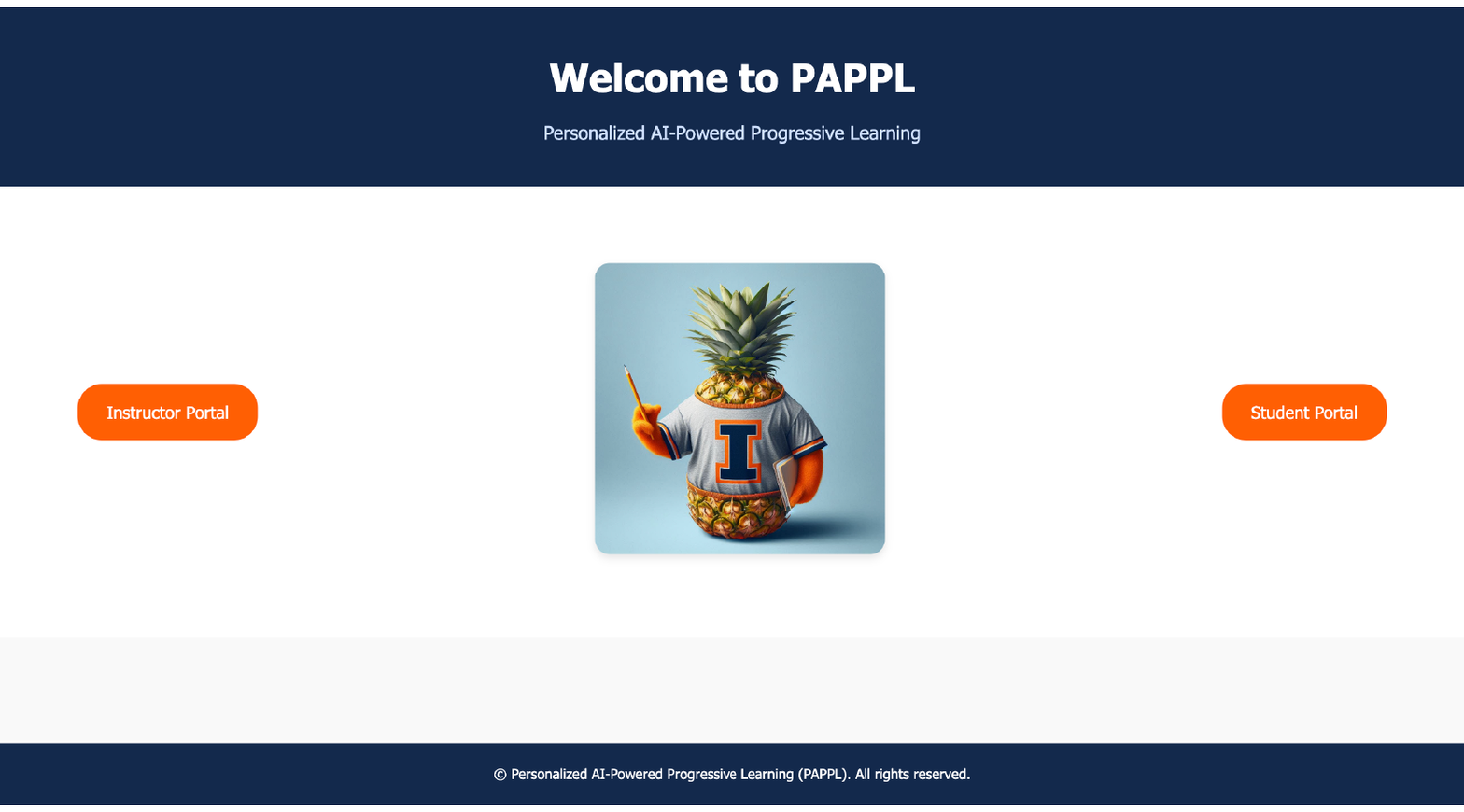}
        \caption{}
        \label{fig:SPsub1}
    \end{subfigure}
    \hfill
    \begin{subfigure}[b]{0.45\textwidth}
        \centering
        \includegraphics[width=0.25\textheight, angle=270]{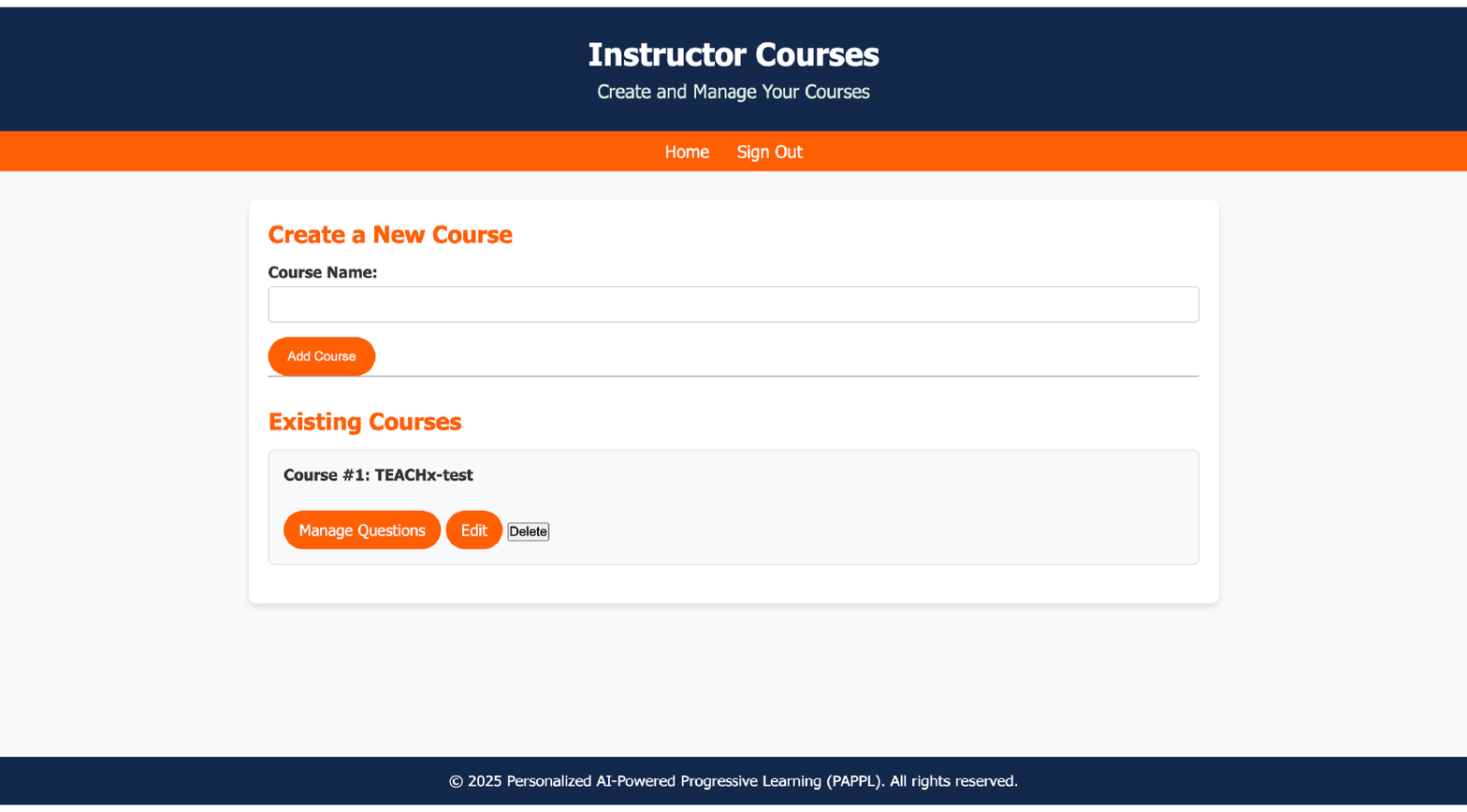}
        \caption{}
        \label{fig:SPsub2}
    \end{subfigure}
    \hfill
    \begin{subfigure}[b]{0.45\textwidth}
        \centering
        \includegraphics[width=0.25\textheight, angle=270]{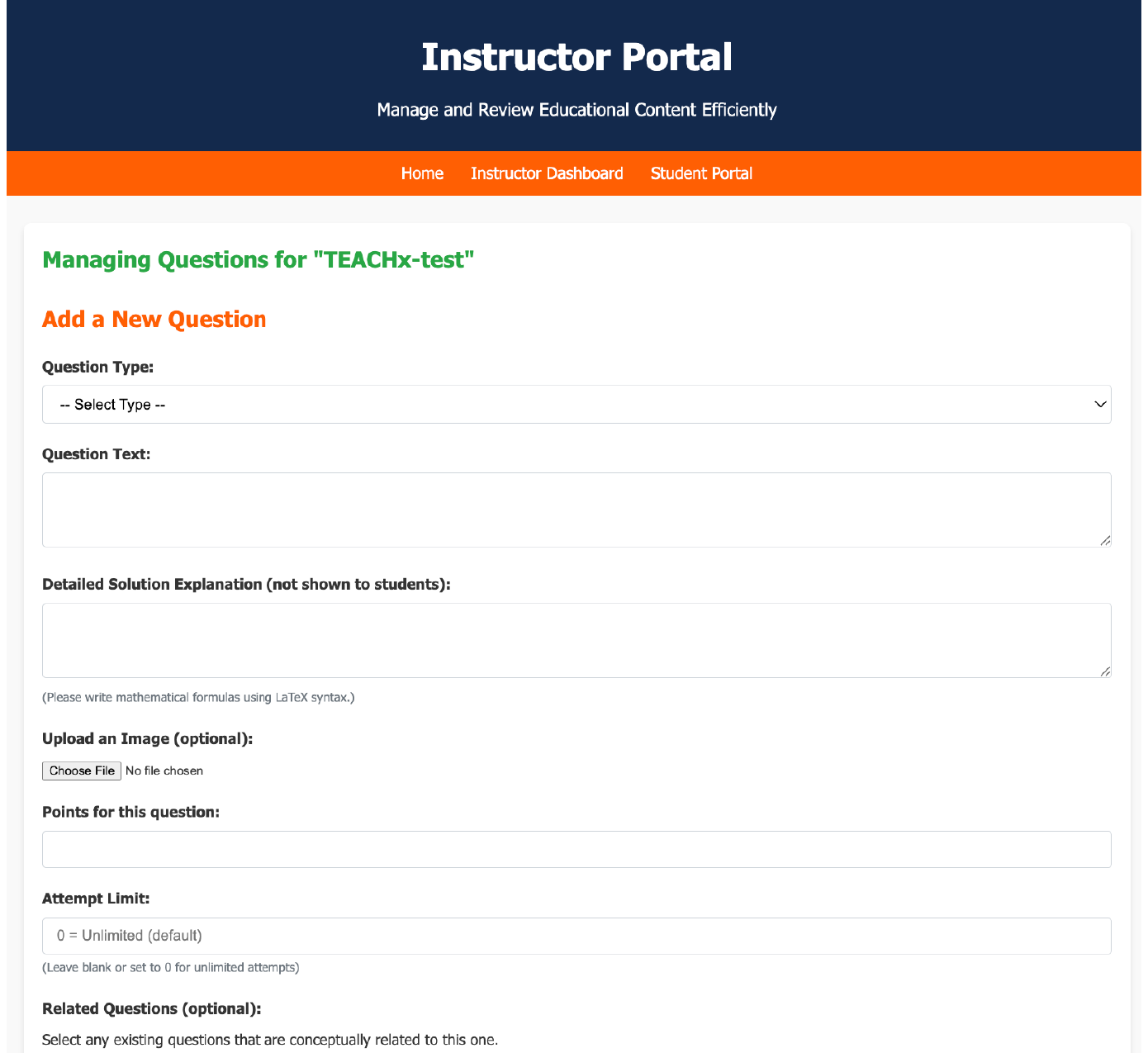}
        \caption{}
        \label{fig:SPsub3}
    \end{subfigure}
    \hfill
    \begin{subfigure}[b]{0.45\textwidth}
        \centering
        \includegraphics[width=0.25\textheight, angle=270]{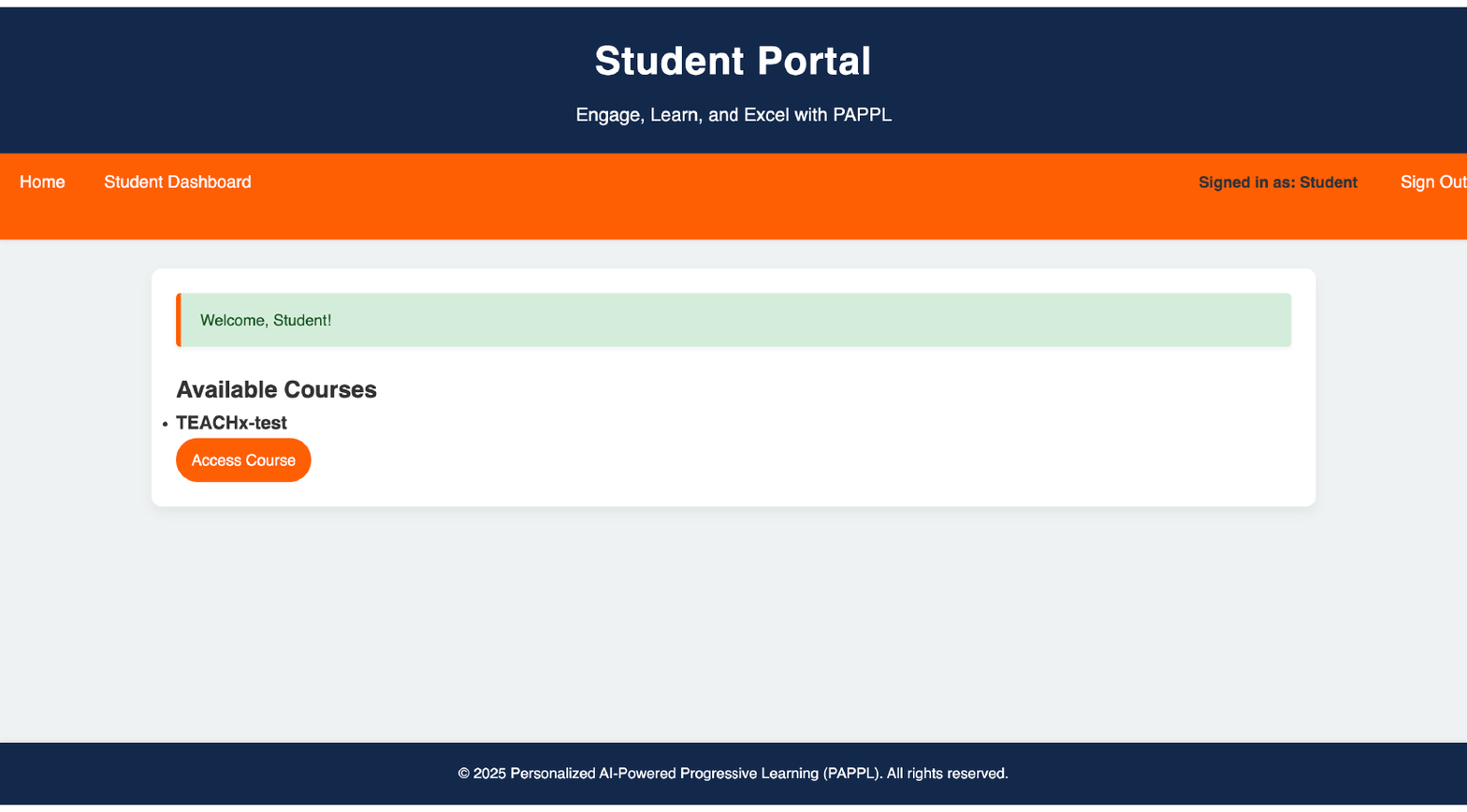}
        \caption{}
        \label{fig:SPsub4}
    \end{subfigure}
    \hfill
    \begin{subfigure}[b]{0.45\textwidth}
        \centering
        \includegraphics[width=0.25\textheight, angle=270]{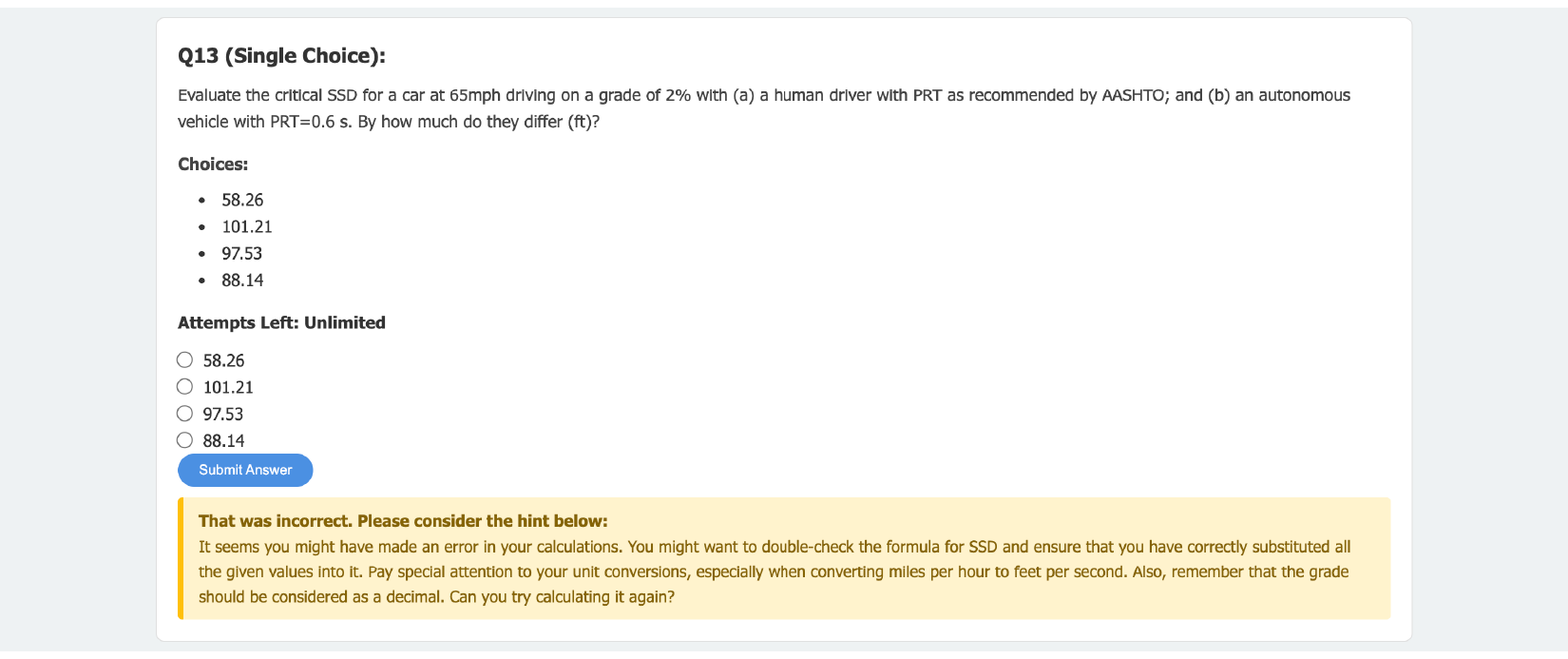}
        \caption{}
        \label{fig:SPsub5}
    \end{subfigure}
    \caption{User Interface for the PAPPL: (a) Home Page, (b) Instructor Dashboard, (c) Question Defining Platform, (d) Student Portal, and (e) System's response to a sample Question from CEE 415 Course: Senior Design for Transportation and Water Resources students at UIUC.}
    \label{fig:interface}
\end{figure}

The platform provides a comprehensive and intuitive web portal for both learners and instructors.  
Students can browse registered courses, attempt questions, and immediately receive AI-generated hints tailored to their recent errors.  
Instructors manage courses and author different types of questions, including multiple-choice, single-choice, true/false, and short-answer, while attaching correct answers, explanatory solutions, contextual descriptions, and optional visual aids (images or plots). Figure~\ref{fig:interface} depicts the user interface for the PAPPL.

\subsection{Backend Infrastructure}
The server logic is implemented in Python. Core design priorities are robustness, extensibility, and low operational overhead, explained thoroughly in the following:

\subsubsection{Content Manager}
All instructional material is stored in a structured repository indexed by course, topic, and sub-topic.  
When an instructor creates or edits a problem, the framework preserves the problem text, associated media, and grading schema.  
This repository enables the dynamic retrieval of appropriate questions and contextual metadata whenever a student submits an answer.
\subsubsection{Learner-State Analyzer}
Behind the scenes, PAPPL securely records every student's attempt, including timestamps, correctness flags, and previously delivered hints.  
By continuously aggregating this history, the analyzer detects recurring misconceptions and builds an evolving learner profile that guides subsequent feedback.
\subsubsection{AI Hint Engine}
Using the learner profile, instructor-supplied context, and the current problem statement, the system constructs context-rich prompts for a GPT-4o model. To be more detailed, the GPT-4o integration constitutes the intelligence core of the platform and is designed in several layers:  
\begin{itemize}
  \item \textbf{Prompt construction}: the back-end aggregates (a) the current item, (b) instructor-provided context, and (c) the learner’s historical errors into a structured prompt that forbids disclosure of solutions.
  \item \textbf{Response handling}: JSON payloads are parsed to distinguish correct verdicts (i.e., for short-answer questions) and to develop explanatory feedback. To mitigate large-language-model hallucination, GPT-4o is invoked with a deliberately low sampling temperature, biasing generation toward higher-probability tokens while permitting limited variation from the context provided by the instructor.
  \item \textbf{Privacy safeguards}: all personally identifying information is stripped before API calls, satisfying FERPA data-protection statutes.
\end{itemize} 
It then returns targeted, personalized hints that steer students toward correct reasoning. Hints become progressively more specific when repeated misconceptions persist, yet remain concise for students who demonstrate academic confidence.

\subsubsection{Feedback \& Analytics Dashboard}
After each interaction, PAPPL updates cumulative statistics and awards points proportional to question difficulty.
Instructors may download an Excel sheet summarizing engagement metrics and analysis: total attempts, correctness ratios, timestamps, and the exact hints provided for each student. These analytics provide adjustment opportunities to teaching strategies and course content.

\subsubsection{Security and Extensibility}
Cross-origin resource sharing (CORS) rules allow secure integration with external front-end clients, and the modular micro-service design facilitates future extensions, such as adopting alternative large-language-model back-ends or additional visualization widgets.

PAPPL modules are also categorized with respect to ITS domains in Table~\ref{tab:pappl_modules}.

\begin{table}[htbp]
\centering
\caption{PAPPL modules mapped to the four classical ITS domains}
\label{tab:pappl_modules}
\renewcommand{\arraystretch}{1.15} 
\begin{tabular}{@{}p{4cm} p{10.5cm}@{}}
\toprule
\textbf{Module} & \textbf{Role in PAPPL} \\
\midrule

\multicolumn{2}{@{}l}{\textit{Expert\,Module}} \\ \midrule
Course Manager   & Create, Read, Update, and Delete (CRUD) interface for courses, persisting metadata through the Course Object-Relational Mapping (ORM) model. \\
Question Bank    & Authoring, editing, and deletion of multiple-choice, single-choice, true/false, and short-answer items; supports image uploads, point values, attempt limits, and links to related questions (Question model). \\ \midrule

\multicolumn{2}{@{}l}{\textit{Student\,Module}} \\ \midrule
Attempt Logger   & Persists every attempt with timestamp, correctness flag, hint text, and awarded points; forms the raw data for learner-state estimation. \\
Learner-State Analyzer & Aggregates attempt history to detect recurring misconceptions and maintain an evolving profile that conditions subsequent feedback. \\ \midrule

\multicolumn{2}{@{}l}{\textit{Tutor\,Module}} \\ \midrule
Hint Generator   & Calls GPT-4o to deliver Socratic hints using prior attempts, related-question context, and instructor solutions. \\
Automatic Grader & For short answers, employs GPT-4o to judge semantic correctness and return a JSON verdict that determines the correctness of the student's response and provides an explanation. \\ \midrule

\multicolumn{2}{@{}l}{\textit{User\,Interface}} \\ \midrule
Student Portal   & Allows learners to browse courses, attempt questions, and receive real-time AI-generated hints. \\
Instructor Dashboard & Web front-end for course and question authoring, plus download of analytics reports. \\ \midrule
\bottomrule
\end{tabular}
\end{table}

\section{Case Study}
\subsection{Study Design}

\begin{figure}[!ht]
  \centering
  \begin{subfigure}[b]{0.45\textwidth}\centering
    \includegraphics[width=0.35\textheight]{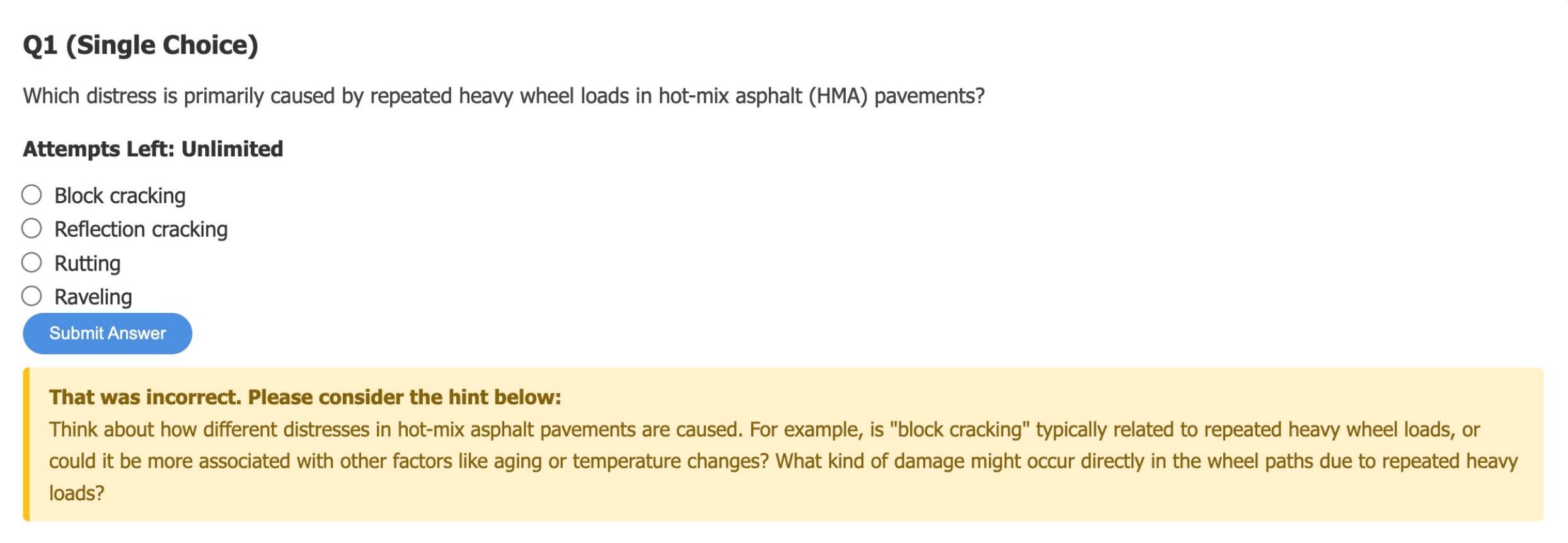}\caption{}\label{fig:testsub1}
  \end{subfigure}\hfill
  \begin{subfigure}[b]{0.45\textwidth}\centering
    \includegraphics[width=0.35\textheight]{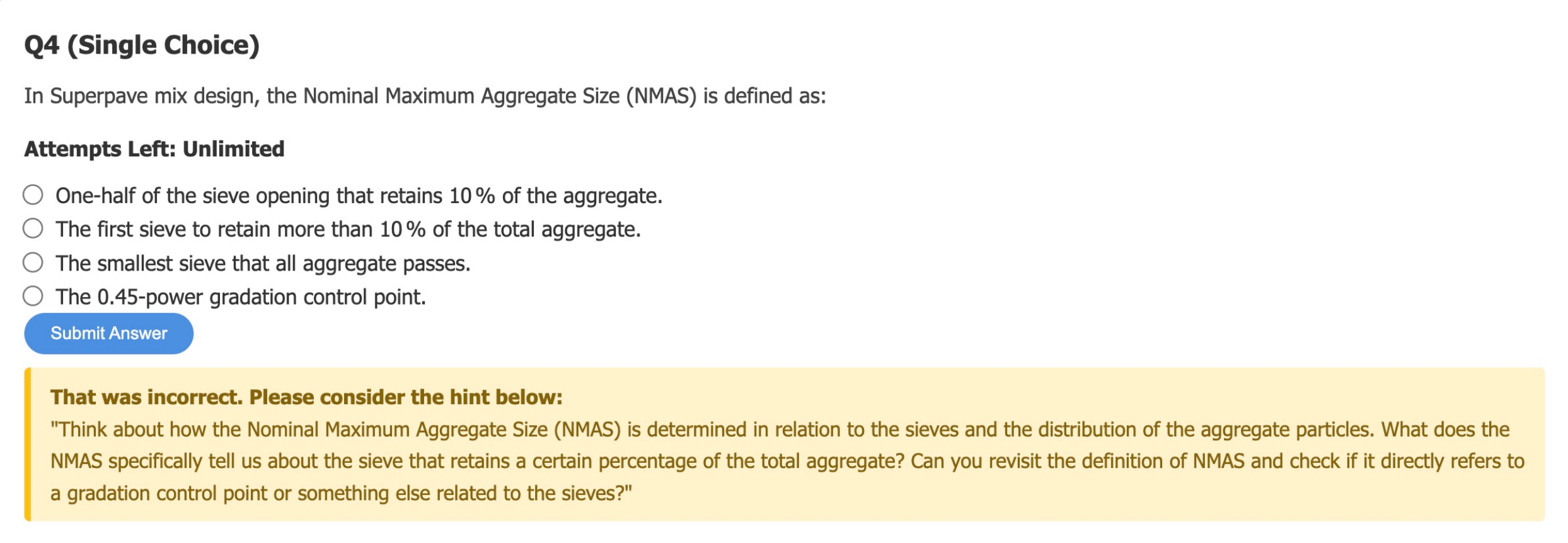}\caption{}\label{fig:testsub2}
  \end{subfigure}\hfill
  \begin{subfigure}[b]{0.45\textwidth}\centering
    \includegraphics[width=0.35\textheight]{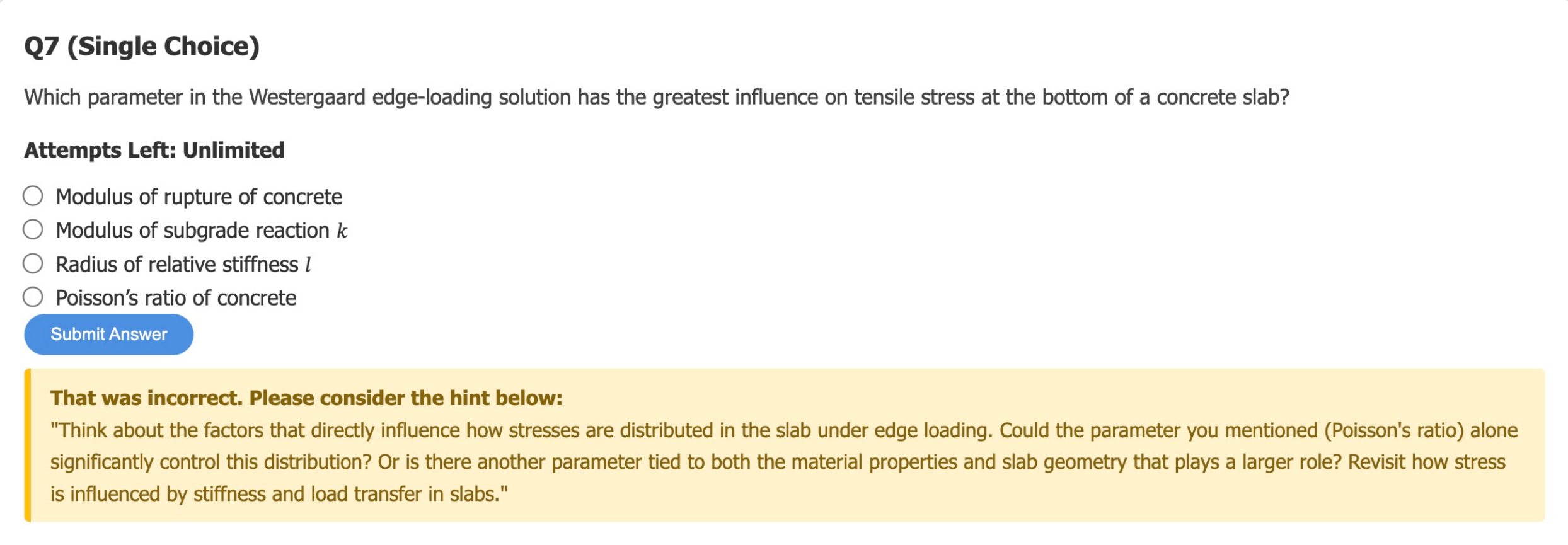}\caption{}\label{fig:testsub3}
  \end{subfigure}\hfill
  \begin{subfigure}[b]{0.45\textwidth}\centering
    \includegraphics[width=0.35\textheight]{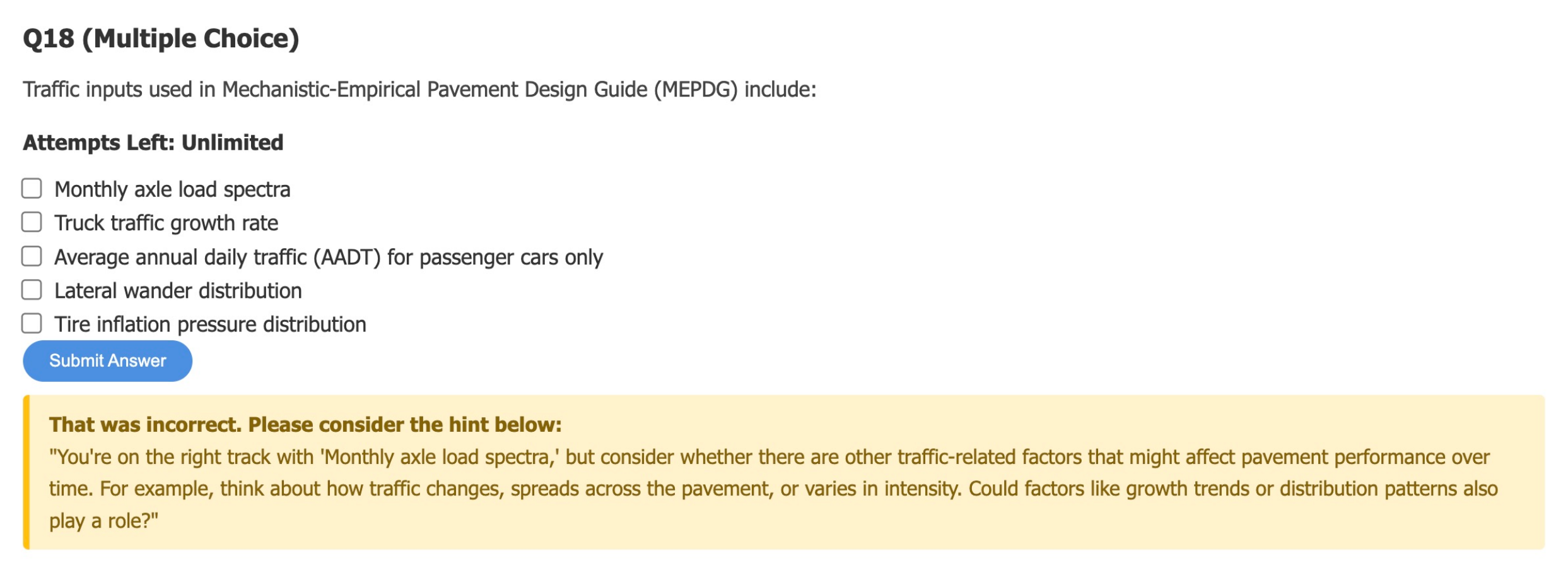}\caption{}\label{fig:testsub4}
  \end{subfigure}\hfill
  \begin{subfigure}[b]{0.45\textwidth}\centering
    \includegraphics[width=0.35\textheight]{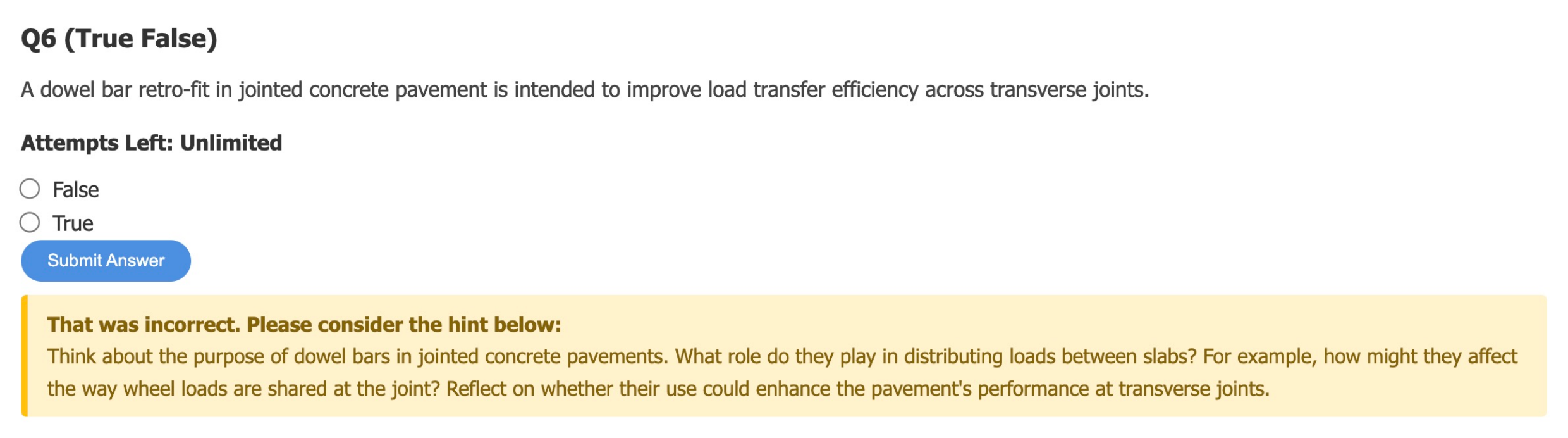}\caption{}\label{fig:testsub5}
  \end{subfigure}
  \caption{PAPPL in action within a STEM-designated field, providing feedback on a pavement-engineering test: (a) surface distresses, (b) material and mix-design fundamentals, (c) structural design methods, (d) traffic loading, and (e) maintenance and rehabilitation treatments.}
  \label{fig:test}
\end{figure}

This experimental study involved graduate students from the transportation group at UIUC and GWU and was structured around four key areas related to pavement engineering, a topic with which the participants were unfamiliar. This provided an excellent opportunity to test the effectiveness of PAPPL in helping students learn a new subject. The study included 25 questions which were designed to cover five areas, each of which is frequently explored in current research: surface distresses~\cite{Arambula_Mercado2016}, material and mix design fundamentals~\cite{Wei2022_SBS_PPA}, structural design methods~\cite{Ceylan2023_CurlingWarping_PhaseII}, traffic loading~\cite{Attia2014}, and maintenance and rehabilitation treatments~\cite{Wilde2014_CostEffective}. A sample question from each area, along with the system's initial response to an incorrect answer, is depicted in Figure~\ref{fig:test}.

The case study aimed to test the usefulness of PAPPL by having participants with a near-zero background in the subject, to assess their learning process in a new domain. Consequently, the participants were divided into two groups: one group used PAPPL and received AI-generated feedback during the assessment, while the other group completed the same assessment without receiving any hints or feedback from the system. These are referred to as the PAPPL and baseline groups, respectively.

To further track the learning progress of the PAPPL group and to evaluate their satisfaction with the overall PAPPL system, they were asked to fill out a questionnaire after completing the pavement engineering test. The full questionnaire is presented in Table~\ref{tab:questionnaire}. The questionnaire consists of 22 items in total, including two open-ended questions. It covers five core dimensions: Effectiveness, Engagement, Adaptivity, Satisfaction, and Accuracy. For the quantitative analysis, we use a 5-point Likert scale (1 = Strongly Disagree to 5 = Strongly Agree). The design intentionally incorporates both negatively worded statements and duplicate questions. Negatively worded questions, such as Q7, Q10, Q16, Q19, and Q20, are included to mitigate the risk of acquiescence response bias, ensuring that participants engage thoughtfully with each question. Furthermore, closely related or duplicate questions, such as Q4 replicating Q3, and Q8 replicating Q6, are used to assess the internal consistency and reliability of participant responses.

\begin{table}[h!]
\centering
\caption{Evaluation dimensions and corresponding questionnaire items for the PAPPL system}
\label{tab:questionnaire}
\begin{tabularx}{\textwidth}{l|X}
\hline
\textbf{Dimension} & \textbf{Questionnaire Items} \\
\hline
\textbf{Effectiveness} &
1. PAPPL's exercises helped me solve pavement-engineering problems. \\
& 2. PAPPL improved my problem-solving skills for pavement-engineering problems. \\
& 3. PAPPL deepened my understanding of pavement-engineering concepts. \\
& 4. Using PAPPL enhanced my grasp of pavement-engineering principles. \\
\hline
\textbf{Engagement} &
5. PAPPL's interactive exercises kept me motivated to solve pavement-engineering problems. \\
& 6. I felt engaged while solving pavement-engineering problems with PAPPL’s AI hints. \\
& 7. I often felt distracted while using PAPPL to solve pavement-engineering problems. \\
& 8. Using PAPPL kept me actively involved in solving pavement-engineering problems. \\
\hline
\textbf{Adaptivity} &
9. PAPPL adapted hints based on my responses and progress. \\
& 10. PAPPL's AI hints were often irrelevant to my specific errors. \\
& 11. PAPPL provided a personalized learning experience tailored to my needs. \\
& 12. PAPPL’s AI hints provided an appropriate level of guidance without revealing answers. \\
\hline
\textbf{Satisfaction} &
13. PAPPL's learning experience was valuable for my pavement-engineering studies. \\
& 14. The PAPPL interface was easy to use and supported my learning. \\
& 15. I would recommend PAPPL to peers studying pavement-engineering. \\
& 16. Using PAPPL was frustrating at times. \\
\hline
\textbf{Accuracy} &
17. PAPPL's AI hints were relevant and helpful for solving pavement-engineering problems. \\
& 18. PAPPL’s AI hints were free of inaccuracies or misleading information. \\
& 19. I often ignored PAPPL’s AI hints because they were not helpful. \\
& 20. PAPPL's AI hints were often repetitive. \\
\hline
\textbf{Open-Ended} &
21. Which PAPPL features best supported your learning? Provide examples. \\
& 22. What challenges did you face with PAPPL, and how could it be improved? \\
\hline
\end{tabularx}
\end{table}

\subsection{Discussion}
\begin{figure}[!ht]
  \centering
  \includegraphics[width=0.9\textwidth]{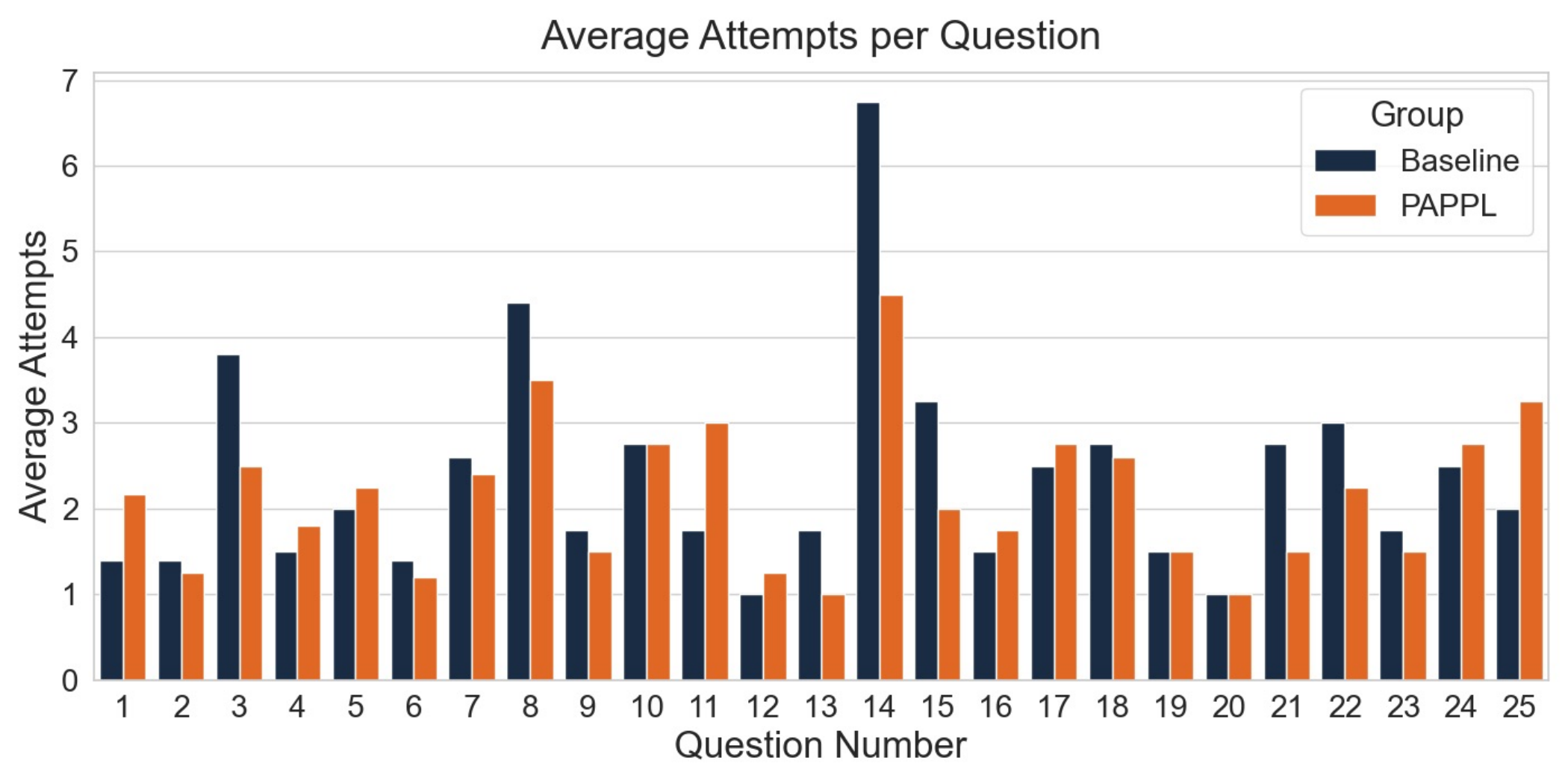}
  \caption{Number of attempts made by each group during the case study.}
  \label{fig:attempt}
\end{figure}

This section summarizes both the quantitative findings from the test and the qualitative findings from the questionnaire gathered from the experiment. Figure~\ref{fig:attempt} presents the average number of attempts required by each group to choose the correct answer for each question. As shown, the PAPPL group, who received AI-generated feedback, was generally able to solve the problems with fewer attempts. Also, our analysis revealed that approximately 30\% of participants in both groups were able to select the correct answer on their first attempts. However, in the second attempt, the percentage of the PAPPL group who succeeded increased to 55\%, while the baseline group showed a more modest improvement, reaching a success rate of 40\% in their second attempt.

\begin{figure}[!ht]
  \centering
  \includegraphics[width=0.9\textwidth]{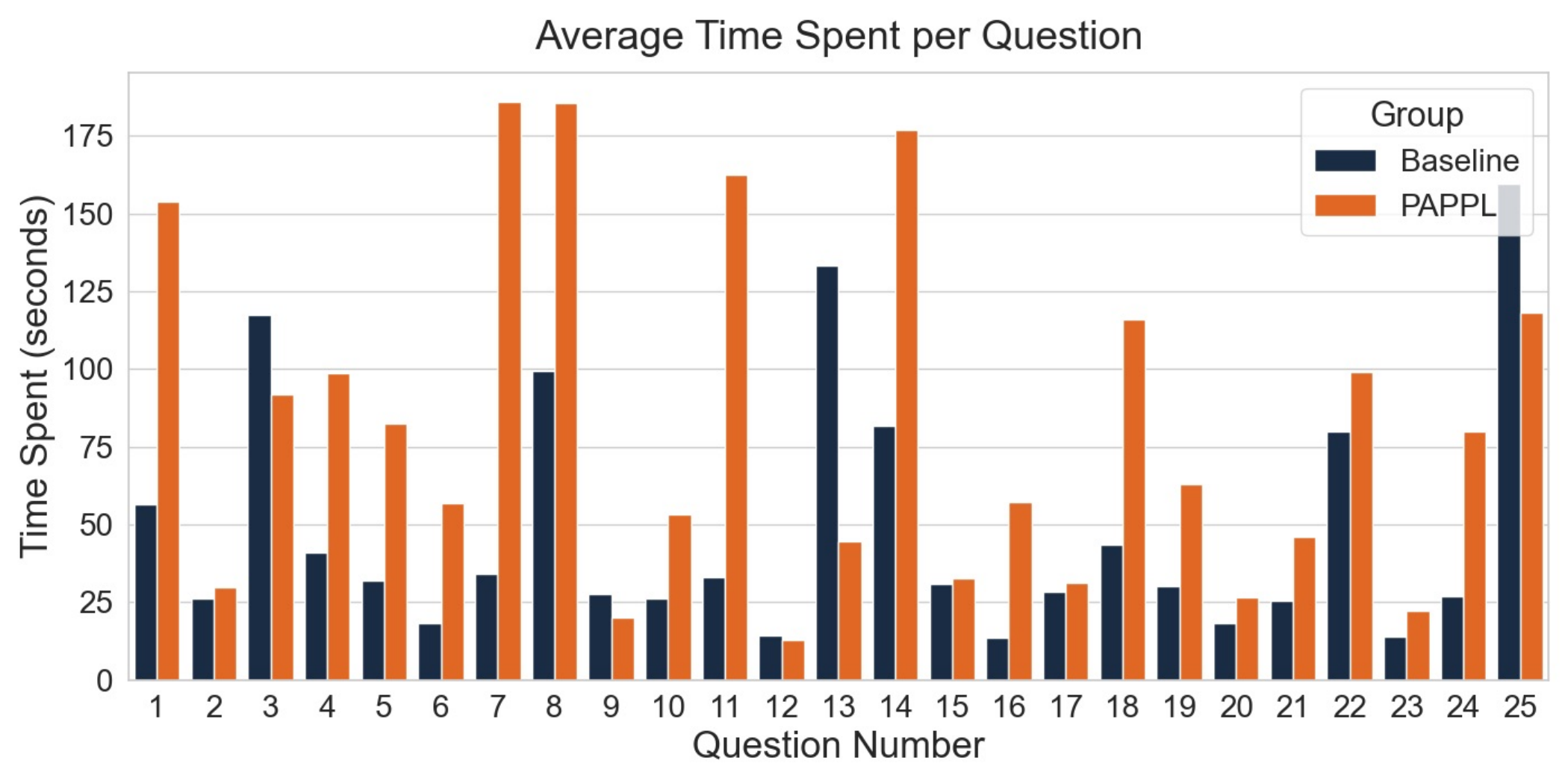}
  \caption{Average time spent per question.}
  \label{fig:time}
\end{figure}

Figure~\ref{fig:time} illustrates the average time spent per question by each group. The PAPPL group spent more time on each question, likely due to reflecting on the problem and reading the AI-generated hints. The combination of increased time investment, fewer required attempts, and a notable improvement in success rate from the first to the second attempt suggests a deeper and more effective learning process compared to the baseline group. These observations are further supported by the questionnaire results discussed in the following.

\begin{figure}[!ht]
  \centering
  \includegraphics[width=0.5\textwidth]{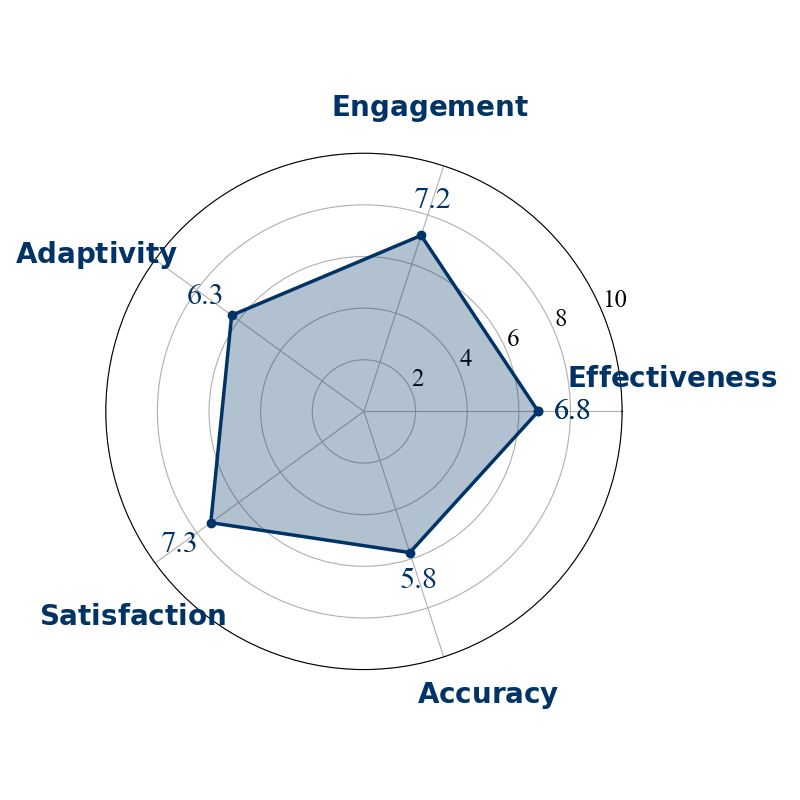}
  \caption{Average user-experience scores across five dimensions (rescaled to a 10-point scale). Higher scores indicate a more positive user experience.}
  \label{fig:qualitative}
\end{figure}

To ensure clear interpretation of the questionnaire, all negatively worded questions were reverse-coded so that higher scores consistently reflect more positive responses across all dimensions. As shown in Figure~\ref{fig:qualitative}, all five user-experience dimensions received average scores above 5, indicating overall satisfaction across all aspects of the PAPPL system. Specifically, the scores were 6.8 for Effectiveness, 7.2 for Engagement, 6.3 for Adaptivity, 7.3 for Satisfaction, and 5.8 for Accuracy. These results suggest that participants in the PAPPL group found the system generally effective, engaging, adaptive, and satisfactory, with moderate confidence in the accuracy of the AI hints. Additionally, all participants passed the duplicate-question consistency check, with no response pair differing by more than one Likert point.

However, there is still room for improvement based on responses to the open-ended questions. Several participants noted that the AI-generated hints were occasionally repetitive or failed to adapt to their chosen answers. In particular, hints for multiple-choice questions were noted to be unclear or unhelpful. To address these issues, further prompt engineering could be employed to enhance the clarity and responsiveness of the hints, thereby improving the overall learning experience. Additionally, while the current sample and single-session design offer valuable early insights, the dataset is not large enough for a definitive assessment of PAPPL’s educational impact. Future deployments across multiple courses will be necessary to gather more comprehensive data.

\section{Conclusion}
This study introduced the PAPPL platform, a powerful tool designed for education at all levels to provide students with more personalized and adaptive educational experiences. It can deliver personalized, AI-driven hints tailored to individual student interactions. By analyzing previous responses and error patterns, PAPPL utilizes a large language model (LLM) as its intelligence core to provide context-specific feedback, significantly enhancing the learning experience. Although the platform is flexible and can utilize any LLM, it defaults to advanced features of GPT-4o to ensure suggestions are relevant and pedagogically sound. This creates a more adaptive and empowering educational environment for students.

Despite these advancements, several gaps remain that future research should address. Detailed student profiles, which PAPPL relies on to comprehensively capture individual learning behaviors, error patterns, and response accuracy, might not exist or be limited at each institution. Such detailed and extensive profiles are crucial as they would offer valuable insights into students' learning processes. Future research should focus on reducing the tutor module's dependency on robust and nuanced student profiles to improve AI capabilities in predicting students' learning patterns. This can further improve students' learning outcomes and create a more engaging and supportive educational environment.

Additionally, prompt engineering techniques require further refinement to fully leverage the inherent capabilities of advanced language models, particularly in visual content interpretation. Given GPT-4o and future models' capacity for image processing, structured prompts enabling AI to accurately interpret visual materials, such as diagrams, graphs, and plots, should be developed to enhance AI-generated hints' precision and educational value, and include visualizations in the hints provided. Furthermore, accessibility remains a critical gap; research should explore alternative interaction methods, notably text-to-speech functionalities, to cater specifically to students with visual impairments or reading disabilities, thereby ensuring equitable educational opportunities. Finally, while initial findings from the case study show promise, further research is needed to systematically evaluate the platform's effectiveness across diverse educational contexts and disciplines. Such evaluations would provide essential insights into the adaptability and scalability of the PAPPL platform, ensuring its evolution into a comprehensive, versatile educational tool capable of significantly enhancing teaching practices.

\section{Acknowledgements}
This work was supported by the National Science Foundation under Grant No. 2047937.

\noindent Authors would also like to thank all the participants in the data collection process from both George Washington University of the University of Illinois at Urbana-Champaign.

\section*{Ethical Statement}

The authors declare that they have no known competing financial interests or personal relationships that could have appeared to influence the work reported in this research.

\bibliographystyle{unsrtnat}
\bibliography{reference} 

\end{document}